# Seeing Topological Order and Band Inversion in Optical Diatomic Chain

Jun Jiang, Zhiwei Guo, Weiwei Zhu, Yang Long, Haitao Jiang, Jie Ren*, and Hong Chen*

MOE Key Laboratory of Advanced Micro-Structured Materials, School of Physics Science and Engineering, Tongji University, Shanghai 200092, China

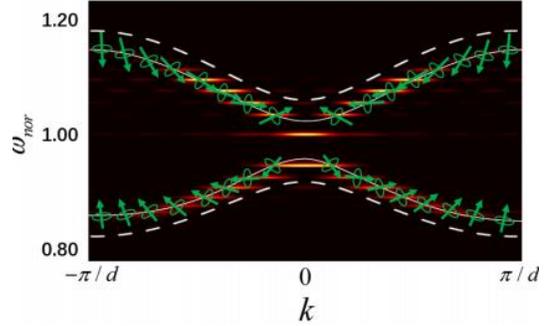

**ABSTRACT:** Recent realizations of exotic topological states in condensed matter and cold atoms have advanced the exploration for topological characteristics, such as invariant topological orders and band inversion. Here we construct a 1D optical diatomic system made of split ring resonators with electric-magnetic couplings to implement the topological features. We experimentally measure the dispersion relationship and sub-lattice pseudo-spin vectors by detecting field strength distributions, which determines the accompanying winding number of bulk band as the topological order. The sub-lattice pseudo-spin vector inversion observed at band-edges evidences the band inversion. We further reconfirm the band-inversion-induced topological phase transition by measuring the symmetry exchanges at band-edges. These results have shown the great potential of split ring resonator platform to experimentally explore the advanced topological characteristics, such as high winding numbers or even high Chern numbers in more exotic setups.

Topology, as a mathematical discipline that studies invariant throughout continuous changing, has attracted much broad attention in physics since the advent of topological insulators (TIs)[1]. Promising applications of TIs and relevant concepts can be found from dissipationless quantum transport[2] to noise-robust quantum computation[3], where the nontrivial properties are characterized by the topological invariant of Bloch bands. The most well-known 2D example is the Chern number[4], an quantized integer related to the Berry's phase[5] crossing the Brillouin zone, which characterizes the fundamental physical phenomena of quantum (spin) Hall effect[6-11] and TIs. For a 1D system, such as SSH model originally introduced in an electronic context named after Su-Schrieffer-Heeger[12], the topological invariant is usually defined as winding number[13]. Though certain topological features related to winding number have been explored in crystals[14] and more recently in various artificial periodic lattices for cold atoms[15], acoustic waves[16] or photons[17], directly measurement of winding number has not yet been implemented in experiments.

Experimentally, the technical difficulties make measuring topological features in such systems remain a challenging issue, which have prompted scientists to explore other controllable platforms for analogous realizations. Moreover, in addition to measuring winding number, finding a simple method to measure topological features is significant. It is then natural to consider to observe the band-inversion and topological edge states. These phenomena, initially discussed in the context of condensed matter physics[18, 19], was subsequently investigated in artificial structures, such as microwave resonator chains[20, 21], a coupled double Peierls chain[22] and quasi-1D cavities[23].

Here we construct a versatile optical platform to reveal topological features with a 1D microwave optical diatomic system, which is made of split ring resonators (SRRs) with hybrid electric-magnetic couplings[24]. We directly measure the winding number as the topological order in this SSH-like diatomic chain to distinguish different topological phases. In particular, we measure the bulk band dispersion and the associate pseudo-spin vectors. The band inversion is demonstrated by detecting the pseudo-spin inversion and symmetry exchange of the states at band-edges. All these results can be obtained by probing the field profile distributing over the whole chain. Although determing topological features has been experimentally explored in cold atom[15, 25], photonic systems[26, 27] and acoustic systems[16, 28] by measuring geometric phase, measuring the topological order in 1D systems, i.e., the so-called winding number, is still open and yet to be solved. In what follows, we will show how to see the winding number and band inversion in our controllable experimental platform.

## ■EXPERIMENTAL SETUP AND ANALYTICAL MODEL.

We employ the diatomic SRR chain with alternating coupling $\kappa_1$ and $\kappa_2$ to directly measure topological order and band inversion. As depicted in Figure 1a, the chain is placed in a plane parallel waveguide whose cutoff frequency exceeds the eigenfrequencies of the system to ensure that the far-field radiation is suppressed and the near-field coupling, a key factor in tight-binding description, is dominant. An example of a unit cell, with the gaps of the two split rings are next to each other (we refer to this configuration as P1), is shown in the partial enlarged drawing in Figure 1a. Starting from configuration P1, the rings can be rotated around their centers simultaneously to mimic polyacetylenes with different topological phases[12] and six representative configurations are obtained and shown in Figure 2. In the following we will consider a general case whose unit cell is formed by a pair of arbitrarily oriented split rings (Figure 1b). The equation of motion can be approximated as follows[30]:

$$-\omega_{nor}^2 \begin{pmatrix} a_k \\ b_k \end{pmatrix} = \begin{pmatrix} -1 & \kappa_1 + \kappa_2 e^{-ikd} \\ \kappa_1 + \kappa_2 e^{ikd} & -1 \end{pmatrix} \begin{pmatrix} a_k \\ b_k \end{pmatrix}, \quad (1)$$

where $\omega_{nor}$ denotes the frequency normalized to the resonant frequency $\omega_0$, $\begin{pmatrix} a_k \\ b_k \end{pmatrix}$ specifies the Bloch function of cell-periodic current states with wavevector $k$. It should be noted that the coupling between two SRRs is always composed of two types of coupling: magnetic and electrical coupling[24, 29], but one of them may be predominant in some typical configurations, When the gaps of two SRRs are next to each other (configuration P1), the dominant coupling is electrical and positive, while when gaps are exactly opposite to each other (configuration P6), the dominant coupling is magnetic and negative.

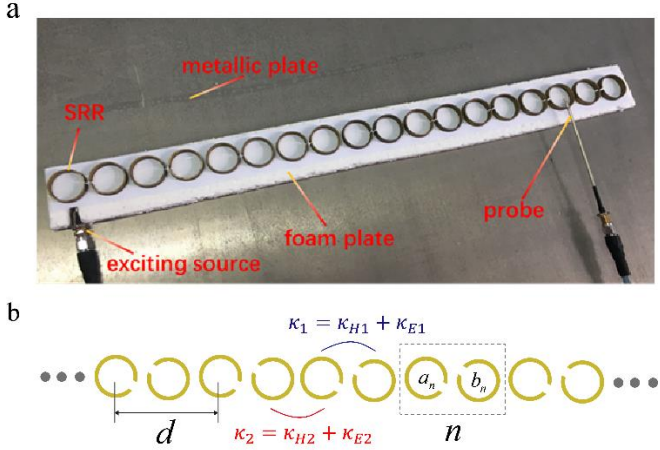

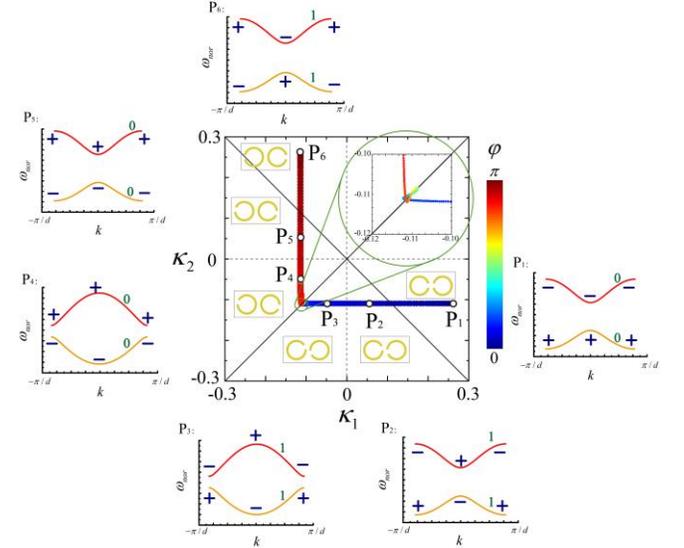

**Figure 1.** 1D optical SRR chain system with dimerized sites. (a) The experimental setup. The chain is composed of identical coupled copper SRRs and put on a one-centimeter-thick EPS foam plate, sandwiched between two metallic plates (note that the top plate is not shown). The two metallic plates are separated by 30mm, forming a plane parallel waveguide whose cutoff frequency is 5.0GHz. The ring dimensions are $r$ =10mm, $h$ =5.0mm, $w$ =1.0mm and $g$ =1.5mm, the corresponding resonant frequency measured in the waveguide is $\omega_0/2\pi$ =1.90GHz. The chain's unit cell length is $d$ =48mm and all rings are separated by the same center distance $d/2$. The probe connected with a network analyzer for data extraction is not shown. (b) Schematic illustration of the optical SRR chain, with A and B sublattices indicated by gap orientations. The total coupling inside the unit cell (between two unit cells) is approximately denoted by $\kappa_1 = \kappa_{H1} + \kappa_{E1}$ ($\kappa_2 = \kappa_{H2} + \kappa_{E2}$), where $\kappa_E$ ($\kappa_H$) are the electrical (magnetic) couplings between two SRRs[21].

## ■TOPOLOGICAL PHASE TRANSITION IN OPTICAL SYSTEM.

In our special system, topological phase transition can be realized by rotating all rings instead of changing their locations like others[30, 31]. As shown in Figure 2, starting from P1 configuration, the trace (represented by color-coded line) of alternating coupling ($\kappa_1$ and $\kappa_2$) is plotted to demonstrate the topological phase transition as rotating all rings anticlockwise from 0 to π simultaneously. Since the bandgap would close at $|\kappa_1|=|\kappa_2|$, once the color-coded line crosses the diagonal lines indicated by $|\kappa_1|=|\kappa_2|$, the bandgap closing and reopening process can be seen, which is a typical characteristic of topological phase transition.

For further making sure that the topological phases change, we pick six diatomic configurations with different rotation radians (rotating phases) $\phi$. These configurations are denoted by P1, P2, P3, P4, P5 and P6 and their corresponding $\phi$ values are 0, 0.18, 0.3, 2.85, 2.97 and π rad. Then we label the winding numbers of bulk bands and symmetries of band-edge states (the even and odd symmetry are represented by symbol '+' and '−') on their corresponding dispersion curves. The winding number of each band is obtained as [30]:

**Figure 2.** Topological phase transition. The color-coded line denotes the evolving trace in $\kappa_1 - \kappa_2$ coupling space as rotating all SRRs simultaneously anti-clockwise from two-gap face to face configuration (P1, phase 0) to two-gap-back-to-back configuration (P6, phase π). The color denotes the rotating phase as indicated by the color bar. Six representative points denoted from P1 to P6 are picked out to illustrate the topological phase transition in detail. Their corresponding unit cells and dispersion curves are plotted, together with the symmetry of band-edge state (here the symbol +/− represents even/odd symmetry) and the winding number of each band (0 or 1). Obviously, configurations P1, P4 and P5 belong to topological trivial class while other three configurations P2, P2 and P6 are topological nontrivial.

$$w_{winding} = \frac{i}{\pi} \int_{-\pi/d}^{\pi/d} \langle u_k|\partial k|u_k\rangle dk = \frac{\theta_k}{2\pi}\bigg|_{k=-\pi/d}^{k=\pi/d}, \quad (2)$$

where the normalized eigenvector $|u_k\rangle = \frac{1}{\sqrt{2}}\begin{pmatrix} e^{-i\theta_k} \\ \mp 1 \end{pmatrix}$ with $\theta_k$ being the determined through $\tan(\theta_k) = \langle\sigma_y\rangle/\langle\sigma_x\rangle = \langle u_k|\sigma_y|u_k\rangle/\langle u_k|\sigma_x|u_k\rangle$ ($\sigma_x$ and $\sigma_y$ are the first and second Pauli matrix). It is obvious that if we track the point with coordinates ($\langle\sigma_x\rangle$, $\langle\sigma_y\rangle$) as evolving the wave vector $k$ through the Brillouin zone, the winding number of the corresponding bulk band can be obtained visually. The determination of symmetry of band-edge state can be readily measured from the current profile in SRRs as the field strength distribution.

From P1 to P2, the inter-intra-couplings change the order from $|\kappa_1|>|\kappa_2|$ to $|\kappa_1|<|\kappa_2|$. As a consequence, the bandgap closes and reopens at the Brillouin zone center and the symmetries of two band-edge states at $k$ = 0 exchange, similar to the band inversion process in electronic systems[32-35]. The topological transition is also indicated by the discrete change of winding numbers between 0 and 1, calculated from Equation (2). From P2 to P3, merely the intracoupling $\kappa_1$ changes sign. As a result, the bands deform and the bandgap location moves from the center to the edge of the Brillouin zone, with their winding numbers intact. From P3 to P4, the color-coded line cross the line of $\kappa_1 = \kappa_2$ backwards and forwards for several times (see the partially enlarged drawing in the inset of Figure. 2). The situations for P4 to P5 and P5 to P6 can be also discussed similarly. When directly comparing P1 and P6, it can be regarded that the topological phase transition is associated with the band inversion happens at k = ±π and the winding numbers switch discretely from 0 to 1.

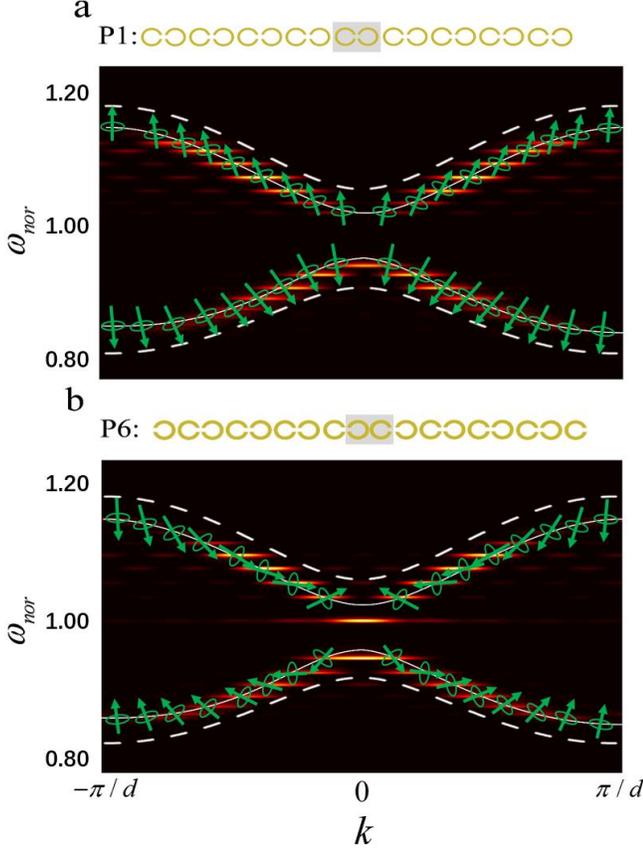

**Figure 3.** Measurement of the dispersions and pseudospin textures accompanying each band. (a) Dispersion diagrams $S(\omega, k)$ for P1 configuration constructed by 18 SRRs. The white dash lines denote the theoretical dispersion curves. The brighter spots are the components of the measured dispersion curves (guided by white solid lines). The sublattice pseudospin vectors are represented by the green arrows which are placed on their corresponding measured eigenmodes with the first phasor is always taken in vertical upward as the reference. (b) Dispersion diagrams $S(\omega, k)$ for P6 configuration, with experimentally measured (white solid lines), theoretically calculated dispersion curves (white dash lines) and sublattice pseudospin vectors. The existence of the midgap state is clearly observed, indicating that P6 configuration belongs to the topological non-trivial class.

## ■ MEASUREMENT OF BAND INVERSION AND WINDING NUMBERS.

The determination of 2D topological order, Chern number, has been theoretically proposed and experimentally implemented in 2D photonic crystals[36, 37], microwave networks[38], cold atoms[39-42]. The 1D topological order, winding number, recently has also been explored experimentally in circuit QED[43,44] and skyrmions systems[45]. Our 1D optical diatom system provides alternative feasibly controllable platform, thus we follow a different scheme to determine the topological order experimentally. The topological order, that is winding number in our system, can be determined by measuring the rotating pseudospin vectors in $k$ space directly.

Here we impose P1 and P6 chain consisting of 18 SRRs to implement the microwave measurements. The lattice sites are occupied by copper SRRs that are coupled by the evanescent near-field electromagnetic field. The tight-binding-like coupling terms κ depend on the relative orientations between two rings. The resonance frequency ω0 of an isolated SRR is around 1.90 GHz and corresponds to the on-site energy of atoms in the tight-binding model. The experiments are performed in the frequency range of 1.50 to 2.35 GHz with a Network Analyzer connected to two probes by 50 Ohm transmission lines (the probes are both non-resonant circular loop antennas of 2mm radius with high impedance). One antenna acting as the exciting source is placed parallel with and directly below the first ring to excite the ring. The magnetic field along the chain is then detected by another probe moving along the axis of the chain at a certain distance above the rings. Though the 50-Ohm loaded probe is not matched to the analyzed structure, the detected magnetic-field signal is proportional to the current in the ring. Thus we can regard the detected signals as the flowing alternating currents.

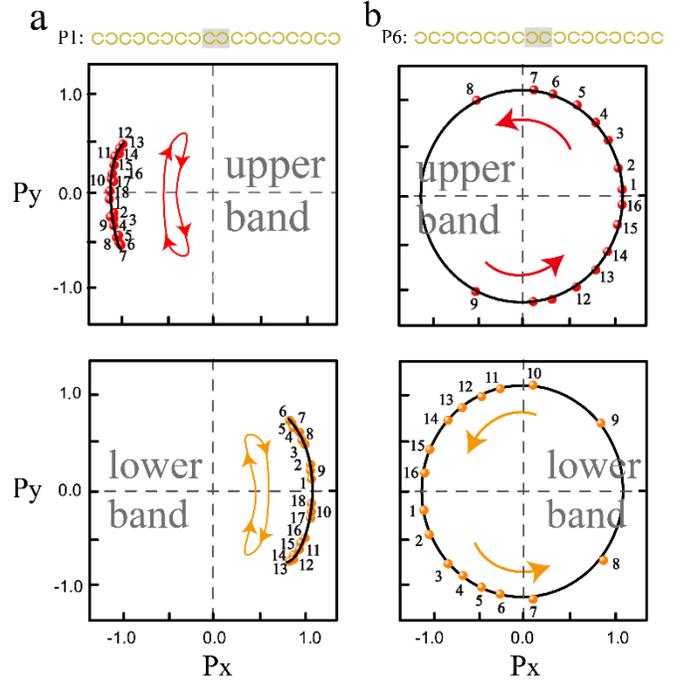

**Figure 4.** Measurement of the winding numbers. (a) Measurement of the components of normalized pseudospin vectors in Px–O–Py plane for the upper and lower bands of P1 configuration. The discrete balls indicate the ends of the measured pseudospin vectors. The solid black lines are the calculated results based on Eq. (5). The solid arrow lines are the eye-guided lines for the evolution path when we tune the quasi-momentum k cross the Brillouin zone from –π to π. Obviously, the winding numbers are both zero. (b) Measurement of pseudospin vectors for upper and lower bands of the topologically nontrivial P6 configuration. The meanings of balls and lines are the same with those of (a) and the winding numbers are both one.

With the measured current in each ring, we can determine the dispersion characteristics from the spectrum function, or say, the dynamical structural factor[46]

$$S(\omega, k) = \left| \sum_{n=1}^{9} \left[ a_{n,\omega} e^{-ikd(n-\frac{1}{2})} + b_{n,\omega} e^{-ikdn} \right] \right|^2, \quad (3)$$

where $a_{n,\omega} = I_{n,A}(\omega) e^{\varphi_{n,A}(\omega)}$, $b_{n,\omega} = I_{n,B}(\omega) e^{\varphi_{n,B}(\omega)}$ and $\varphi_n$, $X(\omega)$ ($X$ is A or B) are amplitude and phase of the current on the SRRs in the nth unit cell. As shown by the measured $S(\omega,k)$ in Figure. 3, the experimental dispersion curves fitted by brighter spots are in good agreement with white dash lines based on calculated eigenfrequencies (details can be found in the part of methods). The discrete wave vectors $k$ are acquired due to the finite short length (unit number) of the system. Nevertheless, the clear existence of midgap state in P6 band structure demonstrates that the chain configuration P6 belongs to the topological non-trivial class, consistent with the theoretical prediction.

Further, by applying discrete Fourier transform to obtain the quasi-periodic Bloch functions:

$$|u_k\rangle = \begin{pmatrix} a_k \\ b_k \end{pmatrix} = \frac{1}{C} \begin{pmatrix} \sum_{n=1}^{9} a_{n,\omega} e^{-ikd(n-\frac{1}{2})} \\ \sum_{n=1}^{9} b_{n,\omega} e^{-ikdn} \end{pmatrix}, \quad (4)$$

with C the normalization factor, we can obtain the two components of the pseudo-spin as [17]:

$$\begin{aligned} P_x &= \langle \sigma_x \rangle = b_k^* a_k + a_k^* b_k = \pm \cos(\theta_k) \\ P_y &= \langle \sigma_y \rangle = i b_k^* a_k - i a_k^* b_k = \pm \sin(\theta_k) \end{aligned}. \quad (5)$$

We make the components of normalized pseudospin vectors in $P_x$–O–$P_y$ plane for the upper and lower bands. As shown in Figure 4, the ordinal numbers are labelled beside the balls, aim to show how the pseudospin vector evolves in $k$ space. With Equation (2), the winding numbers of configuration P1 and P6 are exact 0 and 1 for both ends (upper and lower), respectively. Excellent agreement with the theoretical prediction is clearly seen. Owing to the finiteness of the short chain limited by our experiments, the distribution of the balls is not so even. Even so, it does not influence or mislead our judgement of the winding numbers acquired from distinct topological bands. We further draw all measured pseudospin vectors on the dispersion curves with the first phasor in Figure 3a always taken in vertical upward as the reference. Comparing Figure 3a (Figure 4a) and Figure 3b (Figure 4b), it is quite apparent that the direction of the pseudospin vectors is inverted at the edge of Brillouin zone, corresponding to the band inversion[1].

## MEASUREMENT OF THE SYMMETRY OF THE BAND-EDGE STATES.

Finally, we measure the symmetry of band-edge states as a complementary evidence of the topological phase transition and band inversion[47]. The eigenfrequencies at band-edge are indicated by the black vertical lines in Figures 5a and 5f, which show the amplitudes of currents on the 18 SRRs for P1 and P6 chain. With the point by point measurements, the current profile distributions at band-edge frequencies can be easily obtained. Results are displayed in Figures 5b-e and Figures 5g-j. We can see that for P1 chains, the modes at band-edges share the same symmetry in both lower and upper bands, meaning that the chain is topological trivial[28, 47]. For P6, the modes at band-edges have different symmetries in both lower and upper bands, indicating that the chain must be topological nontrivial[29, 47]. These results agree with the measurement of winding numbers. Combining the dispersion curves of P1 and P6 chain, the measured symmetries of the current profiles at k = ±π are inverted: 1↔8, 4↔5, which also can be found in the even-odd symmetry exchange of current profiles on the central two SRRs.

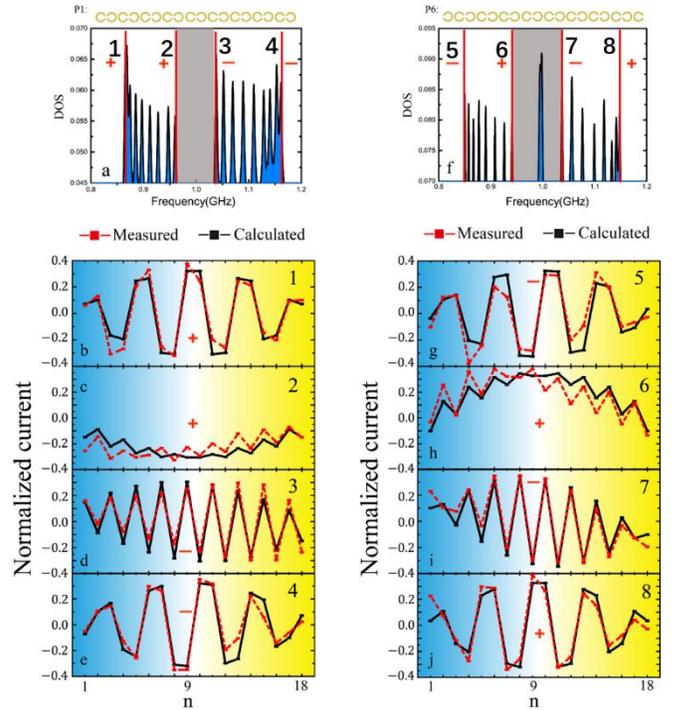

**Figure 5.** Measurement of the symmetry of the band-edge states. (a) Measured current spectrum on the 10th SRR for P1 chain, which is randomly-selected. Measuring on other SRR does not change the current spectrum qualitatively. (b)-(e) Measured current profile as a function of site position for P1 chain at the band-edge frequencies, which are identified by the vertical lines 1-4 in (a). Red circular markers represent experimental results and black square markers represent calculated results. (f) Measured current spectrum for P6 chain. (g)-(j) Measured current spatial distribution for P6 chain at the bandedge frequencies are identified by the vertical lines 5-8 in (f). The markers + and – denote the symmetry and antisymmetry of current profiles, respectively.

## CONCLUSIONS

We have shown that a 1D periodic optical system of SRRs is an excellent platform to realize advanced concepts such as winding number, band inversion and topological midgap states. The experimental samples can be made with minimal cost and the results are intuitively easy to understand. The optical system is rather unique in the sense that Bloch states in real spaces can be measured straightforwardly and, at the same time, the Bloch states depending on wave vector k can be determined with discrete Fourier transform. As such, the global or say non-local properties can be measured locally. These attributes made it possible to straightforwardly determine the winding number of a bulk band, as well as other topology-related properties. Though all the results in this article are demonstrated in a one-dimension system, these methods have the potential to explore the relationship between the symmetry of the eigenmodes, geometrical and topological properties of bulk bands, as well as the bulk-edge correspondence in higher dimensions, which may also be useful to other applications, such as designing more complex topological artificial matters, reconstructing the geometric curvature and Bloch state Hamiltonian, and implementing topological quantum computing[48].

## METHODS

**Identify the symmetry of the band-edge state.** The eigenvectors of Equation (1) for both upper and lower band are expressed as

$$|u_{k,\pm}\rangle = \begin{pmatrix} \frac{\kappa_1 + \kappa_2 \cos(kd) - i\kappa_2 \sin(kd)}{\sqrt{(\kappa_1 + \kappa_2 \cos(kd))^2 + (\kappa_2 \sin(kd))^2}} \\ \mp 1 \end{pmatrix}. \quad (6)$$

Because the coupling coefficient can be positive or negative in our system, we then discuss the symmetries of band-edge states in four cases: 1. $\kappa_1 > 0, \kappa_2 > 0$; 2. $\kappa_1 > 0, \kappa_2 < 0$; 3. $\kappa_1 < 0, \kappa_2 > 0$; 4. $\kappa_1 < 0, \kappa_2 < 0$. For case 1: If $|\kappa_1| > |\kappa_2|$, the chain is topological trivial. At $k = 0$, the wavefunction corresponding to the upper band

$$|u_+(k=0)\rangle = \begin{pmatrix} \frac{\kappa_1 + \kappa_2}{|\kappa_1 + \kappa_2|} \\ -1 \end{pmatrix} = \begin{pmatrix} 1 \\ -1 \end{pmatrix} \text{ is odd symmetry, while the}$$

lower band $|u_-(k=0)\rangle = \begin{pmatrix} \frac{\kappa_1 + \kappa_2}{|\kappa_1 + \kappa_2|} \\ 1 \end{pmatrix} = \begin{pmatrix} 1 \\ 1 \end{pmatrix}$ is even symmetry.

At $k = \pm\pi$, we get:

$$|u_+(k=\pm\pi)\rangle = \begin{pmatrix} \frac{\kappa_1 - \kappa_2}{|\kappa_1 - \kappa_2|} \\ -1 \end{pmatrix} = \begin{pmatrix} 1 \\ -1 \end{pmatrix}, \text{ odd symmetry;}$$

$$|u_-(k=\pm\pi)\rangle = \begin{pmatrix} \frac{\kappa_1 - \kappa_2}{|\kappa_1 - \kappa_2|} \\ 1 \end{pmatrix} = \begin{pmatrix} 1 \\ 1 \end{pmatrix}, \text{ even symmetry.}$$

If $|\kappa_1| < |\kappa_2|$, the chain is topological nontrivial.
At $k = 0$, we get:

$$|u_+(k=0)\rangle = \begin{pmatrix} \frac{\kappa_1 + \kappa_2}{|\kappa_1 + \kappa_2|} \\ -1 \end{pmatrix} = \begin{pmatrix} 1 \\ -1 \end{pmatrix}, \text{ odd symmetry;}$$

$$|u_-(k=0)\rangle = \begin{pmatrix} \frac{\kappa_1 + \kappa_2}{|\kappa_1 + \kappa_2|} \\ 1 \end{pmatrix} = \begin{pmatrix} 1 \\ 1 \end{pmatrix}, \text{ even symmetry.}$$

At $k = \pm\pi$, we get:

$$|u_+(k=\pm\pi)\rangle = \begin{pmatrix} \frac{\kappa_1 - \kappa_2}{|\kappa_1 - \kappa_2|} \\ -1 \end{pmatrix} = \begin{pmatrix} -1 \\ -1 \end{pmatrix}, \text{ even symmetry;}$$

$$|u_-(k=\pm\pi)\rangle = \begin{pmatrix} \frac{\kappa_1 - \kappa_2}{|\kappa_1 - \kappa_2|} \\ 1 \end{pmatrix} = \begin{pmatrix} -1 \\ 1 \end{pmatrix}, \text{ odd symmetry. The other}$$

three cases can be discussed like case 1.

**How to calculate and measure the winding number.** The winding number is the number of loops made by a Bloch state around the equator of the Bloch sphere, as $k$ passes through the Brillouin zone. For our model, it can be expressed as follows:

$$w_{winding} = \frac{i}{\pi} \int_{-\pi/d}^{\pi/d} (a_k^* \partial_k a_k + b_k^* \partial_k b_k), \quad (7)$$

Here, we rewrite Equation (6) as follows:

$$|u_{k,\pm}\rangle = \begin{pmatrix} e^{-i\theta_k} \\ \mp 1 \end{pmatrix}, \quad (8)$$

where $\theta_k$ is determined through $\tan(\theta_k) = \kappa_2 \sin(kd)/[\kappa_1 + \kappa_2 \cos(kd)]$. Combing Equations (7) and (8), we get

$$w_{winding} = \frac{\theta_k}{2\pi} \Big|_{k=-\pi/d}^{k=\pi/d}, \quad (8)$$

By numerical calculation, the winding numbers for both upper and lower band are 1 when $|\kappa_1| < |\kappa_2|$, while they are 0 when $|\kappa_1| > |\kappa_2|$. Though calculating the winding number is simple, measuring winding number is still a big challenge. In the following, we will give a theoretical framework about how to measure winding number. With Equations (1) and (6), we notice that

$$\kappa_1 + \kappa_2 \cos(kd) = (b_k^* a_k + a_k^* b_k)(-\omega_{nor}^2 + 1) = \langle \sigma_x \rangle (-\omega_{nor}^2 + 1)$$
$$\kappa_2 \sin(kd) = (ib_k^* a_k + ia_k^* b_k)(-\omega_{nor}^2 + 1) = \langle \sigma_y \rangle (-\omega_{nor}^2 + 1) \quad (10)$$

where $\sigma_x$ and $\sigma_y$ are the first and second Pauli matrix. So, $\theta_k$ in Equation (8) can be acquired by $\tan\theta_k = \langle \sigma_y \rangle / \langle \sigma_x \rangle$, which means that if we plot these points ($\langle \sigma_x \rangle, \langle \sigma_y \rangle$) at every wave vectors k in the equatorial plane of a Bloch sphere, the winding number of the system can be obtained visually. Usually, we employ $\langle \sigma_x \rangle$ and $\langle \sigma_y \rangle$ to represent the two components of the pseudo-spin vectors of Bloch states and the winding number can be interpreted as the winding of the pseudo-spin vectors. All in all, the determination of the cell-periodic Bloch function of a system play a key role for measuring the winding number. For a finite chain in experiment, the cell-periodic Bloch function can be expressed as:

$$\begin{pmatrix} a_k \\ b_k \end{pmatrix} = \begin{pmatrix} \sum_{n=1} a_{n,\omega} e^{-ikd(n-\frac{1}{2})} \\ \sum_{n=1} b_{n,\omega} e^{-ikdn} \end{pmatrix}. \quad (11)$$

In our experiment, the Bloch mode $\begin{pmatrix} a_n \\ b_n \end{pmatrix}$ for both sublattices can be determined by the directly measure (see methods section in the main text) the amplitude $I_n(\omega)$ and phase $\varphi_n(\omega)$ of currents as a function of frequency and their relationships are

$$\begin{pmatrix} a_n \\ b_n \end{pmatrix} = \begin{pmatrix} I_{n,A} e^{i\varphi_{n,A}} \\ I_{n,B} e^{i\varphi_{n,B}} \end{pmatrix}. \quad (12)$$

**Measurement of spatial current distributions.** In our 1D optical system containing 18 SRRs, the pseudospin vector can be determined by measuring the spatial current distribution at each eigenfrequency. The experiments were performed in the frequency range of 1.5 to 2.35GHz with a Network Analyzer connected to two probes by two 50-Ohm transmission lines (the probes are both non-resonant circular loop antennas of 2mm radius with high impedance). One antenna acted as the exciting source was placed parallel with and directly below the first ring to excite the ring. The current along the chain was then detected by another probe moving along the axis of the chain at a certain distance above the rings. For P6 configuration, the spatial current distribution at two edge modes are not shown. The red squares represent the measured results while the black squares denote the calculated results. They are matching well. With each measured spatial current distribution, the two corresponding pseudospin vectors are labeled in green arrow lines.

■ AUTHOR INFORMATION

**Corresponding Authors**
*Email (J. Ren): Xonics@tongji.edu.cn.
*Email (H. Chen): hongchen@tongji.edu.cn.
**Author Contributions**
J. Jiang, J. Ren, and H. Chen conceived the idea. W. W. Zhu, H. T. Jiang, Y. Long joined discussions on the theoretical analysis. J. Jiang and Z. W. Guo performed the experimental measurements and analysis the experimental data; J. Jiang, Z. W. Guo, J. Ren, and H. Chen prepared the manuscript. All authors fully contribute to the research.
**Notes**
The authors declare no competing financial interest.

■ ACKNOWLEDGEMENTS

The authors would like to thank Dr. Paul Knott from the university of Leeds for fruitful discussions. This work is sponsored by the National Key Research Program of China (No. 2016YFA0301101), by the National Natural Science Foundation of China (NSFC) (No. 61621001), by the Natural Science Foundation of Shanghai (No. 17ZR1443800).